\documentclass[aps,12pt,tightenlines,amsmath,amssymb]{revtex4}
\usepackage{graphicx}
\usepackage{dcolumn}
\usepackage{bm}
\usepackage{epsfig}

\newcommand{\be}{\begin{equation}}\newcommand{\ee}{\end{equation}}
\newcommand{\bea}{\begin{eqnarray}}\newcommand{\eea}{\end{eqnarray}}
\def\Tr{\mathop{\mathrm{Tr}}\nolimits} 
\def\Eqref#1{Eq.~(\ref{#1})}
\def\lsmatrix#1#2{\left(\begin{array}{#1}#2\end{array}\right)}

\newcommand{\labeld}[1]{ }
\def\Id{{\bm 1}}
\def\rhoszz{\rho_{S,00}}\def\rhosoo{\rho_{S,11}}
\def\rhosoz{\rho_{S,10}}\def\rhoszo{\rho_{S,01}}
\def\rhozero{{\rho_S(0)\otimes\rho_A}}
\def\mbu{{\widehat{\bm m}}}\def\sigmab{{\bm \sigma}}
\def\np{{n+1}}
\def\HH{{\cal H}}
\def\rhood{\rho_{OD}}

\def\B{e^{-\beta\omega}}
\def\Bn#1{e^{-\beta {#1} \omega}}

\begin{document}
\title{Violation of the zeroth law of thermodynamics for a non-ergodic interaction}
\author{B. Gaveau}
\affiliation{Laboratoire analyse et physique math\'ematique, 14 avenue F\'elix Faure, 75015 Paris, France}
\email{bernardgaveau@orange.fr}
\author{L. S. Schulman}
\affiliation{Physics Department, Clarkson University, Potsdam, New York 13699-5820, USA}
\email{schulman@clarkson.edu}
\date{\today}
\begin{abstract}
The phenomenon described by our title should surprise no one. What may be surprising though is how easy it is to produce a quantum system with this feature; moreover, that system is one that is often used for the purpose of showing how systems equilibrate. The violation can be variously manifested. In our detailed example, bringing a detuned 2-level system into contact with a monochromatic reservoir does not cause it to relax to the reservoir temperature; rather, the system acquires the reservoir's level-occupation-ratio.
\end{abstract}

\maketitle

\section{Introduction\label{sec:intro}\labeld{sec:intro}}

The zeroth law of thermodynamics is a kind of transitivity. Suppose that if system A is in contact with system B they do not exchange energy. Similarly if A is in contact with another system C, they also do not exchange energy. Then if B and C are brought into contact they too will not exchange energy. As suggested by its position among the laws, this observation about Nature is what permits the other laws of thermodynamics to be discussed.

In this article we present a system that violates the zeroth law. It is a model, and the actual physical systems for which this is (sometimes) an excellent approximation would presumably satisfy the zeroth law on time scales for which the modeling broke down. Nevertheless, there may be experimental time scales for which an apparent violation could be exhibited. In any case, like Arnold's cat map \cite{arnold}, this not-quite-physical system serves to elucidate the foundations of thermodynamics.

The key to the breakdown is the fact that our model system is not ergodic. In classical mechanics ergodicity is usually phrased as the equality of time and phase-space averages. This is equivalent to the non-existence of any constant of the motion except energy (we assume the system is confined so as to fix momentum and angular momentum). In the quantum context, we take ergodicity to mean the absence of non-energy constants, and what lies behind the phenomenon we exhibit is just such a quantity.

A fundamental property of temperature (see for example \cite{reif}) is that objects placed in contact are not in equilibrium unless they have the same temperature. In demonstrating the zeroth law breakdown, we will give an example in which two systems at the \textit{same} temperature, when placed in contact, evolve to \textit{different} temperatures. Call one system ``A'' (which will be thought of as a reservoir) and the other ``S,'' a two-level system. After contact (as described below) they will be at different temperatures. If A is then placed in contact with a second system (say S$'$) that has energy spacing different from that of S, S$'$ will evolve to a third temperature. However, if S and S$'$ are brought in contact, they \textit{will} exchange energy, violating the zeroth law.

For the purpose showing the approach to the non-equilibrium stationary state we will use an intermittent contact. However, having attained that state, reservoir contact becomes arbitrary; it can continue to follow our scheme or it can remain indefinitely coupled, with the density matrix of the system remaining fixed and out of equilibrium. To illustrate subtleties of our contact process we will treat \textit{two} interaction schemes. For both of them the systems evolve to temperatures different from the reservoir, and for the one that does \textit{not} violate the zeroth law the specifics of the contact indeed play a role.

\section{Models and numerical results\label{sec:models}\labeld{sec:models}}

As indicated we work with two interaction models, one for didactic purposes only. The two interactions are commonly called ``rotating wave,'' which will exhibit the zeroth law violation, and ``spin-boson,'' which as stated, is included for purposes of contrast. The Hamiltonians on which we focus are
\be
H=H_{_A}\!+H_{_S}\!+\gamma V ,\qquad H_{_A}\!=\sum_k\omega_ka_k^\dagger a_k,\qquad H_{_S}\!=Es^\dagger s
,
\ee
where
\be
V=\sum_k\left( g_ka_ks^\dagger+g_k^* a_k^\dagger s \right) ,\
 \hbox{rotating wave (RW)},
 \label{eq:RW} \\
\ee
or
\be
V=\left(s^\dagger+  s \right)\sum_k g_k \left(a_k+ a_k^\dagger \right) ,
              \ \hbox{spin-boson (SB)}
\,. \label{eq:SB}
\ee
\labeld{eq:RW}\labeld{eq:SB}
Our notation is standard. System A is a collection of harmonic oscillators (thought of as photons or phonons) and $a^\dagger$ and $a$ are its creation and annihilation operators. For system S we have similar operators, $s^\dagger$ and $s$, although since we take S to have only two levels, squares of these operators are zero. The quantity $\gamma$ multiplying the potential is a convenient overall scaling. Both interactions are physically significant. For RW, using the Wigner-Weisskopf method it is shown in \cite{scully} that if the oscillator states have a thermal distribution with inverse temperature $\beta$ \cite{note:units}, then the ratio of probabilities for the excited and ground states of S is $e^{-\beta E}$---the usual equilibrium. RW dynamics has, besides $H$, the constant of motion
\be
N\equiv \sum_k a_k^\dagger a_k+ s^\dagger s        \qquad \hbox{(excitation number)}
\,,
\label{eq:excitationconserve}
\ee
\labeld{eq:excitationconserve}
and is not ergodic. From the demonstration \cite{scully} it would seem that there are no thermodynamic consequences.

This satisfying situation is the result of approximations and assumptions on A\@. Consider a \textit{single} frequency reservoir that is detuned from S, i.e., $E\neq\omega$\@. We want A to act as a reservoir. It begins at temperature $1/\beta$ and couples to S for a short time, after which they are decoupled; A is refreshed and they are re-coupled. Iterate. This models the Markovian assumption in studies of dissipative dynamics in that A loses its correlations with S\@. Similarly, in diffusion a particle connects and disconnects in successive encounters with other particles (viewed as a reservoir), with correlations carried away by translational degrees of freedom. Our coupling-decoupling has the following mathematical realization: let the initial density matrix be $\rho(0) = \rho_S(0)\otimes \rho_A(0)$ with $\rho_A(0)=\left(1-e^{-\beta\omega}\right)\exp\left(-\beta H_{A}\right)$\@. Evolving under $H$, $\rho(\Delta t)=U\rho(0)U^\dagger$, with $U=\exp\left(-iH\Delta t\right)$\@. Next separate the systems and form the reduced density matrix for S, $\rho_S(\Delta t)=\Tr_A\rho(\Delta t)$\@. The nature of A as a reservoir is manifested by assuming that it recovers from its interaction with S, and is again in equilibrium at temperature-$1/\beta$\@. Repeat this many times. Thus the time-$2\Delta t$ state of S is given by $\rho_S(2\Delta t) = \Tr_A U\left\{\rho_S(\Delta t)\otimes \rho_A(0)\right\} U^\dagger$\@. A physical realization could be the passage S through a sequence of sharply tuned cavities---but not tuned to the frequency of S\@. The coupling is weak, but one expects S to eventually acquire the temperature of A; \textit{it does not}. We find, both numerically and analytically, that the relative probability of being excited is $\exp(-\beta\omega)$, \textit{not} $\exp(-\beta E)$\@.

\smallskip

\noindent \textsf{Remark}: We speak of S's disequilibrium in terms of temperature for convenience only. For a 3- (or more) level system there is also a failure to reach the appropriate Boltzmann distribution. Moreover, if one uses the rotating wave coupling, then the zeroth law breakdown described in this article does obtain. In particular, S adopts the level-occupation probabilities of the reservoir and it takes no energy to maintain this disequilibrium. However, only if S's levels are equally spaced can one still speak of temperature. (A similar result in different circumstances was found in \cite{decoherenceequilibration}.)

The disequilibrium, however surprising, is not by itself a consequence of non-ergodicity. The SB model, without the constancy of $N$, behaves similarly, although the final temperature depends on other variables. The intermittent interaction means that the overall effective Hamiltonian is time-dependent, hence need not conserve energy. Thus when contact is broken there can be interaction energy, $\gamma\Tr(\rho(\Delta t) V)$, whose elimination requires external work and thus maintains the disequilibrium~\cite{note:freecoupling}. Where the RW model is different is that---as we will show---in the non-equilibrium stationary state \textit{there is no \hbox{\rm A-S} decoupling energy}. Separating the systems costs nothing and nevertheless disequilibrium is maintained. Since energy is conserved during the contact interval, the coupling energy equals (up to a sign) the change in $\langle H_A+H_S\rangle$\@. Thus if S has reached its steady state, the cost of maintaining disequilibrium is the change in $\langle H_A\rangle$. \textit{For RW this cost is zero}. After quantitative analyses, we will give a qualitative argument for RW's peculiarity, with the role of non-ergodicity evident.

\begin{figure}
\includegraphics[width=0.45\textwidth]{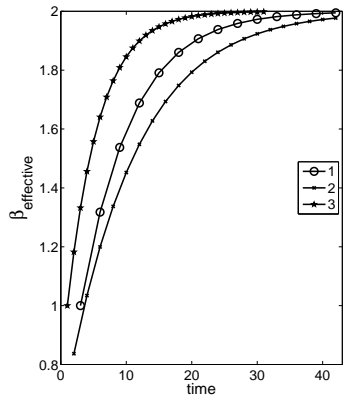}
\caption{Effective-$\beta$ on successive contacts with the $\beta=1$ reservoir, for the RW interaction. Initial conditions as well as duration and strength of the interaction vary. For Run-2 $\rho_S(0)$ is not diagonal. In all cases the same non-equilibrium limit is attained. \label{fig:betaeffective}\labeld{fig:betaeffective}}
\end{figure}

\begin{figure}
\includegraphics[width=0.41\textwidth]{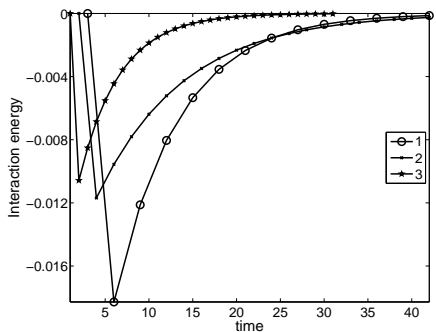}
\caption{Interaction energy at the end of the A-S contact period for the RW interaction. Coupling parameters and initial conditions vary. For Run-2 $\rho_S(0)$ is not diagonal. \label{fig:interactionenergy}\labeld{fig:interactionenergy}}
\end{figure}

In Fig.\ \ref{fig:betaeffective} is shown the time-dependence of $\beta$-effective, computed from the occupancy ratio, for a 2-state RW system for which $E=\omega/2$\@. The initial $\beta$ for S is 1, so that as they repeatedly touch and pull away, S \textit{cools}. Fig.\ \ref{fig:interactionenergy} shows the interaction energy. As S approaches its (non-equilibrium) steady state, the RW interaction energy goes to zero~\cite{note:kurizki}.

In the sequel we first present a density matrix formalism in which the standard equilibration result is shown to be approximate. Then we specialize to single-frequency reservoirs, where the evidence is unambiguous: For RW, S adopts the \textit{occupation ratio} of the bath, not its temperature. If A has frequency $\omega$, that ratio is $\exp(-\beta\omega)$, giving S an effective temperature $T_\mathrm{eff}=(E/\omega) T_\mathrm{bath}$\@. This is true whether the A-S contact duration is long or short and whether or not S begins in a diagonal density matrix. For SB, S tends to an effective temperature that is (in general) not that of the bath. For a certain limit of $\gamma$ and $\Delta t$ the temperature of S tends to infinity. Moreover, for RW maintaining disequilibrium costs \textit{no} work, i.e., the coupling and decoupling requires no energy once S has become stationary. However, for SB, in its nonequilibrium steady state, during each decoupling the bath gains energy, an increase that comes at the expense of whatever external process implements the decoupling (cf.\ \cite{ratchet}).

\section{Small time calculation\label{sec:smalltime}\labeld{sec:smalltime}}

For a general reservoir, single or multiple frequency, we begin with $\rho(0)=\rho_S(0)\otimes\rho_A$, where $\rho_S(0)$ is arbitrary and $\rho_A=e^{-\beta H_{_A}}/Z_A$, $Z_A=\Tr e^{-\beta H_{_A}}$\@. For small $\Delta t$
\be
\rho(\Delta t)=\rho(0)-i \Delta t[H,\rho(0)]-\frac{\Delta t^2}{2}[H,[H,\rho(0)]]+\hbox{O}(\Delta t^3)
\,.
\label{eq:bg3}
\ee
\labeld{eq:bg3}
In analyzing \Eqref{eq:bg3} we use the following: 1)~$[H_A,\rho_S(0)\otimes\rho_A]=0$; 2)~the diagonal matrix elements of $V$ on the eigenvectors of $H_A$ are zero (true for both RW and SB). After several steps \Eqref{eq:bg3} reduces to
\bea
\rho_S(\Delta t)&=&\rho_S(0)-i\Delta t[H_S,\rho_S(0)]   \nonumber\\
      &&\quad
        -\frac{\Delta t^2}{2} [H_S,[H_S,\rho_S(0)]]
        \nonumber\\
       &&\quad -\frac{\gamma^2 \Delta t^2}{2} \Tr_A [V,[V,\rhozero]]+\hbox{O}(\Delta t^3)
\,.   \label{eq:bg6}
\eea
\labeld{eq:bg6}
Although our results are more general, we specialize \Eqref{eq:bg6} to a ``diffusion limit'' \cite{gardiner}, namely, $\gamma\to\infty$ and $\Delta t\to0$ such that
\be
\bar\gamma\equiv\gamma\sqrt{\Delta t}\to\hbox{const}
\,.
\label{eq:gammabar}
\ee
\labeld{eq:gammabar}
In this limit we compute $\left[\rho_A(\Delta t)-\rho_A(0)\right]/\Delta t$, leading to
\be
\frac{d\rho_S}{dt} = -i[H_S,\rho_S(0)]
               -\frac{\bar\gamma^2}{2} \Tr_A [V,[V,\rho_S\otimes\rho_A]]
\nonumber\label{eq:bg8}
\,.
\ee
\labeld{eq:bg8}

\begin{figure}
\includegraphics[width=0.475\textwidth]{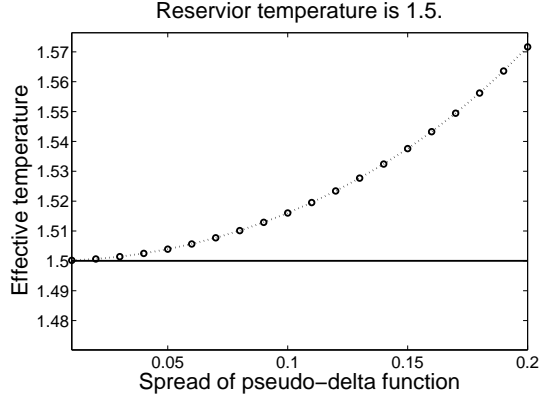}
\caption{Variation of temperature as the spread of a Gaussian pseudo-$\delta$-function is varied.  Temperature, $T$, for S is computed from $(1-\rhoszz)/\rhoszz=\exp(-E/T)$ with $\rhoszz$ given in \Eqref{eq:bg30}\@. (For a Lorentzian pseudo-$\delta$-function the deviation is larger.) \label{fig:integrationspread}\labeld{fig:integrationspread}}
\end{figure}

We first apply these results to RW. Calculating the appropriate matrix elements and commutators, we obtain two independent equations. For the diagonal
\be
\frac{d\rhoszz}{dt}=-\bar\gamma^2\sum_k
      \frac{ \left|g_k\right|^2}{1-e^{-\beta\omega_k}}
        \left[
         e^{-\beta\omega_k} \rhoszz
         - \rhosoo
         \right]
\label{eq:bg24}
\,.
\ee
\labeld{eq:bg24}
The off-diagonal element satisfies
\be
\frac{d\rhosoz}{dt}=
\left[iE-\frac{\bar\gamma^2}{2}\sum_k\left|g_k\right|^2
                        \frac{1+e^{-\beta\omega_k}}{1-e^{-\beta\omega_k}}
\right]\rhosoz
\ee
(and $\rhoszo={\rhosoz}^*$). As a result $|\rhoszo|\sim e^{(iE-\Gamma)t}\to0$, where
\be
\Gamma=\frac{\bar\gamma^2}{2}\sum_k\left|g_k\right|^2
                        \coth\left(\frac{\beta\omega_k}{2}\right)
\nonumber\,.\ee
Using $\rhosoo=1-\rhoszz$ in \Eqref{eq:bg24} yields for the limit
\be
\rhoszz(\infty)=\frac{    \sum_k\left|g_k\right|^2
                   \coth\left(\frac{\beta\omega_k}{2}\right)\frac1{1+e^{-\beta\omega_k}}
                   }
                   {
                       \sum_k\left|g_k\right|^2
                   \coth\left(\frac{\beta\omega_k}{2}\right)
                   }
\label{eq:bg30}
\,.
\ee
\labeld{eq:bg30}
If $g_k$ is sharply peaked at $k_0$ such that $E=\omega_{k_0}$, a short calculation shows that
\be
\frac{\rhosoo(\infty)}{\rhoszz(\infty)}\sim e^{-\beta\omega_{k_0}}=e^{-\beta E}
\,.
\ee
This recovers thermal equilibrium, and is the result of Einstein \cite{einstein}. However, this conclusion depends on the $\{g_k\}$\@. \textit{In general, one does \textbf{not} recover thermal equilibrium.} Fig.\ \ref{fig:integrationspread} shows how effective temperature varies with the width of a Gaussian approximation to a $\delta$-function. The temperature is off, albeit not grossly.

For the monochromatic, detuned thermal bath, ``$g_k$'' is a $\delta$-function at some $k$, but now $E\neq \omega_{k}$. \Eqref{eq:bg30} yields the probability ratio $e^{-\beta\omega_{k}}$, the dropoff exponential of the bath. As a result S has effective temperature
\be
T_\mathrm{eff}=T_\mathrm{bath}\frac {E~}{\omega_{k_0}}
\quad\left(\hbox{or}\quad
\beta_\mathrm{eff} = \beta\frac{\omega_{k_0}}E \right)
\label{eq:RWtemperatures}
\,.
\ee
\labeld{eq:RWtemperatures}

We turn to the SB interaction. Calculating the matrix elements in the limit (\ref{eq:gammabar}),  one obtains
\bea
\frac{d\rhoszz}{dt}&=&-\bar\gamma^2
        \left( \sum_k\left|g_k\right|^2
        \coth\frac{\beta\omega_k}{2}
                \right)
        \left[ \rhoszz
        - \rhosoo
         \right]
\nonumber \\
\frac{d\rhosoz}{dt}&=& iE \rhoszo
  -\bar\gamma^2 \sum_k\left|g_k\right|^2
                        \coth\frac{\beta\omega_k}{2}
                \left[\rhoszo-\rhosoz\right]
\nonumber
\eea
It follows that $\rhoszz\to\frac12$ and $\rhoszo\to0$ for $t\to\infty$\@. The system tends to infinite temperature, which is consistent with~\cite{ratchet}, although the recoupling scheme differs.

\section{Rotating wave, arbitrary times\label{sec:longtime}\labeld{sec:longtime}}

Returning to RW, we show that \Eqref{eq:RWtemperatures} holds for any interaction duration. The single frequency reservoir only has transitions within $\HH_n\equiv\{|n,0\rangle,|n-1,1\rangle\}$ (1$^\mathrm{st}$ label A, 2$^\mathrm{nd}$ S), for $n\geq1$, a manifestation of non-ergodicity. Taking $g$ real (for simplicity) and $\gamma=1$, the Hamiltonian is
\bea
H_n&=&
      \lsmatrix{lll}{
                 n\omega    &~  & g\sqrt{n} \\
                 g\sqrt{n}  &~  & (n-1)\omega+E }           \nonumber \\
       &=&\left(n\omega+\textstyle{\frac{\omega+E}{2}}\right)\Id+\sigma_z\left(
       \textstyle{\frac{\omega-E}{2}}
       \right)
    + g\sqrt{n}\sigma_x
,
\label{eq:Htwomatrix}
\eea
\labeld{eq:Htwomatrix}
with the $\sigma$'s the Pauli spin matrices. Defining
$\Omega_n\equiv\sqrt{\left((\omega-E)/2\right)^2+ng^2}$,\
$\mbu_n \equiv \left(g\sqrt{n},~0,~(\omega-E)/2\right)/\Omega_n$, and
$\Omega_0\equiv n\omega+(\omega+E)/2$,
the propagator in $\HH_n$ is
\be
e^{-iH_n \Delta t}=
     e^{-i\Omega_0\Delta t}\left[\Id \cos\Omega_n \Delta t - i\sigmab\cdot\mbu_n\sin\Omega_n \Delta t
     \right]
\,.\label{eq:propagatortwo}
\ee
\labeld{eq:propagatortwo}
Assume that $\rho_S(0)$ is diagonal in the energy basis, so that transitions away from each $|n_A,n_S\rangle$ can be analyzed separately (the general case is treated in Appendix \ref{sec:general}). From \Eqref{eq:propagatortwo}, if one begins in S's ground state, the probability of excitation is
$
r_n =
\left({ng^2} / {\Omega_n^2}\right)
\sin^2\Omega_n \Delta t \quad(\leq1)
$\@.
For de-excitation the probability is the same, as is evident from \Eqref{eq:propagatortwo}. Calling the ground state probability, summed over all $n_A$, $q$, the change in $q$ after a single interaction interval is
\be
\Delta q= -q\sum p_n r_n+(1-q)\sum p_n r_{n+1}
\,.
\label{eq:Deltaq}
\ee
\labeld{eq:Deltaq}
In the second sum $r$ takes the index $(n+1)$ because these contributions to $q$ come from $\HH_{n+1}$\@. Their weight in $\rho_A$ is nevertheless $p_n$\@. Both sums can be taken from 0 to $\infty$ since $r_0=0$. The second sum can be written $\sum p_n r_{n+1}=e^{-\beta\omega}\sum p_{n+1} r_{n+1}$\@. Define $\bar\Gamma\equiv\sum p_n r_n$\@. Then \Eqref{eq:Deltaq} becomes
\be
\Delta q= -q\bar\Gamma+(1-q)e^{-\beta\omega}\bar\Gamma
\,.
\ee
Setting $\Delta q=0$ to find the stationary state gives $(1-q)/q=e^{-\beta\omega}$. (It is also easy to see that convergence to this value is exponentially rapid.)

The same argument can be applied to changes in A's probability distribution during the course of S's coupling. The possible transitions and the probability transfer for each are as follows: Losses in $p_n$: [$|n,1\rangle\to|n+1,0\rangle$, $r_{n+1}$],  [$|n,0\rangle\to|n-1,1\rangle$, $r_{n}$]. Gains in $p_n$: $|n+1,0\rangle\to|n,1\rangle$, $r_{n+1}$], [$|n-1,1\rangle\to|n,0\rangle$, $r_{n}$]. Thus
$
\Delta p_n =-\left[ p_n(1-q)r_{n+1} +p_n q r_n\right]
  +\left[p_{n+1}qr_{n+1}+p_{n-1}(1-q)r_n\right]
$\@.
The limiting value of $q$ is given by $(1-q)/q=e^{-\beta \omega}$. Using the fact that initially $p_{n+1}/p_{n}=e^{-\beta \omega}$, we find that $\Delta p_n =0$\@. When S has reached its non-equilibrium steady state there is no change in the reservoir's probability distribution, hence no change in its energy. Moreover, by conservation of energy during the period of contact, there is also no residual coupling energy. (This is also true for a non-diagonal $\rho_S(0)$. See Appendix~\ref{sec:general}.)

\section{Spin-boson: disequilibrium takes work.\label{sec:spinboson}\labeld{sec:spinboson}}

\begin{figure}
\includegraphics[width=0.475\textwidth]{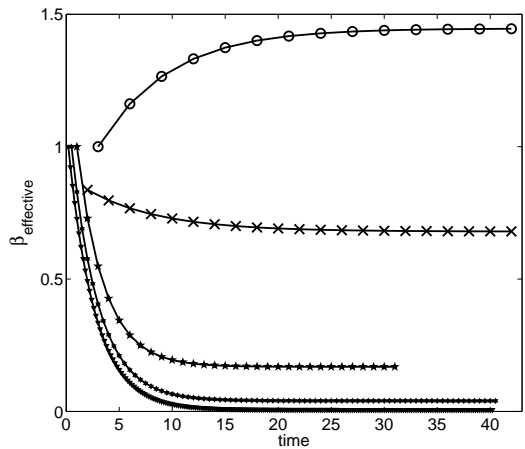}
\caption{Effective-$\beta$ on successive contacts with the $\beta=1$ reservoir. Duration and coupling strength of the SB interaction vary. For the curve using ``$\times$'' symbols $\rho_S(0)$ is not diagonal. The two lowest curves are for decreasing times and fixed $\bar\gamma$, showing the approach to $\beta_\mathrm{effective}=0$\@. Numerically, for small $\Delta t$, $\beta_\infty\approx c\Delta t^2$, with $c\approx0.15$\@. \label{fig:betaeffectiveSB}\labeld{fig:betaeffectiveSB}}
\end{figure}

For SB, numerically, the temperature does converge, but to a value that depends on details (see Fig.\ \ref{fig:betaeffectiveSB}). In the diffusion limit, the propagator becomes $U=\exp(-i\bar\gamma V\sqrt{\Delta t})$\@. The SB interaction induces the following transitions: for $n\geq0$, $(n,\epsilon)\to (n+1,1-\epsilon)$, $\epsilon=0$ or $1$, with matrix element $\bar\gamma\sqrt{\Delta  t}\sqrt{n+1}$, and for $n\geq1$, $(n,\epsilon)\to (n-1,1-\epsilon)$ with matrix element $\bar\gamma\sqrt{\Delta  t}\sqrt{n}$\@. As in \Eqref{eq:Deltaq}, we monitor changes in $q$\@. Now both sums have the same range and using the squares of the matrix elements, $\Delta q=\bar\gamma^2\Delta t \left\{-q\sum p_n \left[n+(n+1)\right]+(1-q)\sum p_n\left[n+(n+1)\right]\right\}$\@. Requiring $\Delta q=0$ implies $q=1-q$, or equal probability of excited and ground states, namely infinite temperature. However, unlike the RW case, the reservoir does not go to its original state at the end of each connection interval. We have calculated, both analytically and numerically, that the energy cost in this limit is $\bar\gamma^2\Delta t$, so that per unit \textit{physical} time there is a cost, $\bar\gamma^2$, to maintain the disequilibrium.

\section{Discussion\label{sec:discussion}\labeld{sec:discussion}}

For RW, we provided a simple argument for the equality of A and S occupation ratios. But even in our ``simple'' argument it may not have been clear where non-ergodicity entered. In the following we do it in words. When S is in its steady state each contact with A must leave it with as much energy as it had before. This does not imply (cf.\ SB) that the final energy of A is unchanged. However, with the additional conservation law of total excitation number (\Eqref{eq:excitationconserve}) for RW, the reservoir must \textit{also} be left with its original excitation. Energy constancy for the system implies constancy for the reservoir. (This argument also applies if S has more than two levels.)

With a multi-frequency reservoir this argument fails, since energy transfer can be accomplished by superpositions of excitations of many frequencies. Moreover, for any finite number of reservoir oscillators greater than one, the decoupling process \textit{does} take energy (see Appendix~\ref{sec:energyflow}).

For both models the systems are in contact with a monochromatic reservoir and have temperatures different from that of the reservoir. By itself this violates nothing: the connection-disconnection process allows a transfer of energy that can maintain disequilibrium. Where the surprise lurks is that for the rotating wave model, \textit{this transfer is zero.} The systems are in intermittent contact, but S is in a state of disequilibrium that takes \textit{no} energy to maintain. Note too that once stationarity has been attained the contact can be intermittent \textit{or not}, since our effect is independent of $\Delta t$\@. Another way of phrasing the violation is that after each contact period it could happen (and does happen for SB) that the final reservoir state has changed its energy in such a way as to maintain the disequilibrium. However, for RW this ``change'' is zero.

In principle, the effect we have reported here must disappear on a sufficiently long time scale, since in the real world the rotating wave interaction is only an approximation. However, if the detuning is slight it will be a good approximation, so that a variation of the physical setup mentioned earlier (an effectively two-state system passing through a succession of detuned cavities) could yield the disequilibrium discussed here.

\begin{acknowledgments}
We are grateful to the Max Planck Institute for the Physics of Complex Systems, Dresden, for its gracious hospitality.
\end{acknowledgments}

\appendix

\section{Exact computation of density matrix evolution, rotating wave (RW) model \label{sec:general}\labeld{sec:general}}

$\HH_n$ is the subspace spanned by $\{|n,0\rangle, |n-1,1\rangle\}$, $n\geq1$, 1$^\mathrm{st}$ label A, 2$^\mathrm{nd}$ label S\@. Because of the non-ergodicity of RW dynamics, transitions can only occur within a given $\HH_n$ and one can consider the following restricted Hamiltonian (taking $g$ real and $\gamma=1$):
\bea
H_n&=&
      \lsmatrix{lll}{
                 n\omega    &~  & g\sqrt{n} \\
                 g\sqrt{n}  &~  & (n-1)\omega+E }           \nonumber \\
       &=&\left(n\omega+\frac{\omega+E}{2}\right)\Id+\sigma_z\left(\frac{\omega-E}{2}\right)
    + g\sqrt{n}\sigma_x
\,,
\label{eq:sm010}
\eea
\labeld{eq:sm010} \\
Define $\Omega_n\equiv\sqrt{\left((\omega-E)/2\right)^2+ng^2}$, $\mbu_n \equiv \left(g\sqrt{n},~0,~ (\omega-E)/2\right)/\Omega_n$, and $\Omega_0\equiv n\omega+\frac{\omega+E}2$. The propagator within $\HH_n$ is
\bea
U_n(t)&\equiv&\exp\left(-iH_nt\right)=
     \exp(-i\Omega_0t)\left[\Id \cos\Omega_n t - i\sigmab\cdot\mbu_n\sin\Omega_n t
     \right]
\nonumber \\
&=&e^{-i\Omega_0t}\lsmatrix{cc}
                {\cos\Omega_n t -i\frac{\omega-E}{2\Omega_n}\sin \Omega_n t
                                      & -i\frac{g\sqrt n}{\Omega_n}\sin\Omega_n t\\
              -i\frac{g\sqrt n}{\Omega_n}\sin\Omega_n t
                                  &  \cos\Omega_n t+i\frac{\omega-E}{2\Omega_n}\sin\Omega_n t}
\nonumber \\
&\equiv&
   e^{-i\Omega_0t}\lsmatrix{cc}
                {c_n -i\delta_n s_n
                                      & -i\gamma_n s_n\\
              -i\gamma_n s_n
                                  &  c_n +i\delta_n s_n}
\,.\label{eq:sm020}
\eea
\labeld{eq:sm020}
The last line implicitly defines $c_n$, etc.

The general form of the initial density matrix is
\be
\rho(0)=\sum_{n\geq0}p_n \left[ \mu\pi_{n,0}+(1-\mu)\pi_{n,1}
                      \right]
                      +\rhood
\,,
\label{eq:sm030}
\ee
\labeld{eq:sm030}
where $p_n=(1-\B)\Bn n$, $\pi_{n_A,n_S}\equiv|n_A\rangle\langle n_A| |n_S\rangle\langle n_S|$, $n_A=0,1,2,\dots$, $n_S=0,1$\@. $\mu$ is real and $0\leq \mu\leq1$\@. The term $\rhood$ is the off-diagonal (for S) contribution. It is given by
\be
\rhood(0)= \left\{\,\zeta\, |0_S\rangle\langle 1_S| +\zeta^* |1_S\rangle\langle 0_S|\right\}
                     \sum_{n\geq0}p_n |n\rangle\langle n|
 \,,\quad \hbox{with~}
                     |\zeta|^2\leq\frac12-\left(\mu-\frac12\right)^2
\,.
\ee
The condition on $\zeta$ is guarantees that $\rho(0)$ is a bona fide density matrix.

The time-dependence of $\rho$ is obtained from $\rho(t)=U\rho(0)U^\dagger$. This can be evaluated for each term in the sum and gives the following for $\rho(t)$:
\be
\rho(t)=p_0\mu\pi_{00}+\sum_{n\geq1}p_n  \mu U_n\pi_{n,0}U_n^\dagger
       +\sum_{n\geq0}(1-\mu)U_{n+1}\pi_{n,1}U_{n+1}^\dagger
       +U\rhood(0)U^\dagger
\,.
\label{eq:sm040}
\ee
\labeld{eq:sm040}
where $U_n$ is the matrix given in \Eqref{eq:sm020} (or its equivalent in bra-ket notation). In \Eqref{eq:sm040} $U_{n+1}$ appears in the second sum, since this acts in ${\cal H}_{n+1}$\@. Note too that ${\cal H}_{0}$ (with basis $|0,0\rangle$) is constant in time. The off diagonal term is
\bea
&&\mskip -12mu\rhood(t)=U\rhood(0)U^\dagger
    \nonumber\\
   && \mskip-8mu
   \ =\left[ \pi_{00}+\sum_{n\geq1}U_n\left(\pi_{n,0}+\pi_{n-1,1}\right) \right]
                \left(     \sum_{\nu\geq0}p_\nu |\nu\rangle\langle \nu|\right)
                           \zeta\, |0_S\rangle\langle 1_S|
                     \left[ \pi_{00}+\sum_{m\geq1}\left(\pi_{m,0}+\pi_{m-1,1}\right) U_m^\dagger  \right]
                     \nonumber\\
   && \qquad                     +\ \hbox{adjoint}
\,,
\label{eq:sm050}
\eea
\labeld{eq:sm050}
This becomes
\be
\rhood(t)=\zeta\sum_{n\geq0}p_n U_n |n,0\rangle \langle n,1| U_\np^\dagger     +\ \hbox{adjoint}
\,.
\label{eq:sm051}
\ee
\labeld{eq:sm051}
In each subspace $\HH_n$, S's operators can be written as 2-by-2 matrices
\be
|0\rangle\langle 0|=\lsmatrix{cc}{1&0\\ 0&0},\
|1\rangle\langle 1|=\lsmatrix{cc}{0&0\\ 0&1},\
|0\rangle \langle 1|=\lsmatrix{cc}{0&1\\ 0&0},\
|1\rangle \langle 0|=\lsmatrix{cc}{0&0\\ 1&0}.
\ee
We evaluate the time evolution. First the projection operators:
\bea
U_n\pi_{n,0}U_n^\dagger&=&
     \lsmatrix{cc}{c_n^2+\delta_n^2s_n^2& i\gamma_n s_n(c_n-i\delta_n s_n)\\
                   -i\gamma_n s_n(c_n+i\delta_n s_n) & \gamma_n^2 s_n^2}
                   \nonumber \\
                   &\equiv&
                   \lsmatrix{cc}{x_n&z_n\\ z_n^*&1-x_n}
                   \nonumber \\
      &=&x_n\pi_{n,0} + (1-x_n)\pi_{n-1,1}+ \left(z_nR_n+\hbox{adjoint}\right)
\,,
\label{eq:sm055}
\eea
\labeld{eq:sm055}
with $R_n\equiv|n,0\rangle\langle n-1,1|$, the raising (with respect to A) operator. Similarly
\bea
U_{n+1}\pi_{n,1}U_{n+1}^\dagger&=&
     \lsmatrix{cc}{ \gamma_\np^2 s_\np^2  & -i\gamma_\np s_\np(c_\np-i\delta_\np s_\np)\\
                   i\gamma_\np s_\np(c_\np+i\delta_\np s_\np) & c_\np^2+\delta_\np^2s_\np^2}
                   \nonumber\\
                   &=&
                   \lsmatrix{cc}{1-x_\np&-z_\np\\ -z_\np^*&x_\np}
                   \nonumber \\
      &=&(1-x_\np)\pi_{\np,0} + x_\np\pi_{n,1}- \left(z_\np R_\np+\hbox{adjoint}\right)
\,.
\label{eq:sm056}
\eea
\labeld{eq:sm055}
The off-diagonal term mixes $\HH_n$'s. We look at
\be
\rhood(t)=\zeta\sum_{n\geq0}p_n U_n   |n,0\rangle  \left(U_\np|n,1\rangle\right)^\dagger
  +\hbox{adjoint}
\,.
\ee
We write this in full:
\bea
\rhood(t)&=&\zeta\sum_{n\geq0}p_n
\left[U_n(1,1)|n,0\rangle+U_n(2,1)|n-1,1\rangle\right]
     \left[U_\np(1,2)|\np,0\rangle+U_\np(2,2)|n,1\rangle\right]^\dagger
\nonumber\\
  &&\quad +\hbox{adjoint}
\nonumber\\
&=& \zeta\sum_{n\geq0}p_n  \left\{
 U_n(1,1)|n,0\rangle  \langle \np,0| U_\np(1,2)^*
+U_n(1,1)|n,0\rangle   \langle n,1|  U_\np(2,2)^*     \right.
\nonumber\\
&& \quad  \left.
+U_n(2,1)|n-1,1\rangle  \langle \np,0| U_\np(1,2)^*
+U_n(2,1)|n-1,1\rangle   \langle n,1|  U_\np(2,2)^*
\right\}
\nonumber\\
&&\quad                                    +\hbox{adjoint}
\label{eq:sm060}
\,.
\eea
\labeld{eq:sm060}
When performing $\Tr_A$, terms with different $n_A$ values go away. Substituting the relevant quantities from the expression for $U_n$ one gets
\bea
\rhood(t)&=& \zeta\sum_{n\geq0}p_n
\left[  (c_n -i\delta_n s_n)  |n,0\rangle   \langle n,1|  (c_\np -i\delta_\np s_\np)
\right]
  +\hbox{adjoint}
  \nonumber\\
  &&\qquad + [\hbox{terms that disappear under~} \Tr_A]
\,.
\label{eq:sm070}
\eea
\labeld{eq:sm070}
Substituting all this into \Eqref{eq:sm040} yields
\bea
\rho(t)&=&p_0\mu\pi_{00}+\sum_{n\geq1}p_n  \mu
      \left\{    x_n\pi_{n,0} + (1-x_n)\pi_{n-1,1}+ \left(z_nR_n+\hbox{adjoint}\right) \right\}
          \nonumber \\
      &&\quad +\sum_{n\geq0}p_n(1-\mu)
    \left\{(1-x_\np)\pi_{\np,0} + x_\np\pi_{n,1}- \left(z_\np R_\np+\hbox{adjoint}\right) \right\}
\nonumber \\ && \quad
+\rhood(t)
\,.
\label{eq:sm080}
\eea
\labeld{eq:sm080}
with $\rhood(t)$ given in \Eqref{eq:sm070}. Changing dummy indices on some of the sums leads to
\bea
\rho(t)&=&p_0\mu\pi_{00}
     +\sum_{n\geq1}p_n  \mu x_n\pi_{n,0}
      +  \sum_{n\geq0}p_\np  \mu  (1-x_\np)\pi_{n,1}
      +\sum_{n\geq0}p_\np  \mu \left(z_\np R_\np+\hbox{adjoint}\right)
          \nonumber \\
      &&\quad
      +\sum_{n\geq1}p_{n-1}(1-\mu)
       (1-x_n)\pi_{n,0}   +\sum_{n\geq0}p_n(1-\mu) x_\np\pi_{n,1}
       \nonumber\\
   &&\qquad \qquad
    -  \sum_{n\geq0}p_n(1-\mu)\left(z_\np R_\np+\hbox{adjoint}\right) +\rhood(t)
\,.
\label{eq:sm090}
\eea
\labeld{eq:sm090}
Our primary interest is $\rho_S(t)$ so we consider what happens when $\Tr_A$ is applied:
\bea
\rho_S(t)&=&|0\rangle\langle0| \left[
p_0\mu+\sum_{n\geq1}p_n  \mu x_n +\sum_{n\geq1}p_{n-1}(1-\mu)(1-x_n)
\right]
\nonumber \\ &&\quad
+|1\rangle\langle1| \left[
  \sum_{n\geq0}p_\np  \mu  (1-x_\np) +\sum_{n\geq0}p_n(1-\mu) x_\np
\right]
\nonumber \\ &&\quad
 +\left\{
 \zeta\sum_{n\geq0}p_n
   \left[ (c_n -i\delta_n s_n) |0\rangle   \langle 1| (c_\np -i\delta_\np s_\np)
   \right]
          +\hbox{adjoint}
 \right\}
\,.
\label{eq:sm100}
\eea
\labeld{eq:sm100}
Now although mathematically $c_n$ and $c_\np$ can simultaneously equal one, physically it is not reasonable, so that generically $|(c_n -i\delta_n s_n)  (c_\np -i\delta_\np s_\np)|<1$. This implies that the off-diagonal terms drop away. We also use the fact that $p_{n+1}=\B p_n$\@.  Without the off-diagonal terms, \Eqref{eq:sm100} becomes
\bea
\rho_S(t)&=&|0\rangle\langle0| \left[
p_0\mu+\sum_{n\geq1}p_n \left[ \mu x_n +e^{\beta\omega} (1-\mu)(1-x_n)
\right]\right]
\nonumber \\ &&\quad
+|1\rangle\langle1| \left[
\sum_{n\geq1}p_n\left[
   \mu  (1-x_n) + e^{\beta\omega} (1-\mu) x_n
\right]\right]
\,.
\label{eq:sm110}
\eea
\labeld{eq:sm110}

We check the $\Delta q$ calculation of the main text. \Eqref{eq:sm110} gives the new value of ``$q$'' as
\be
q'=p_0q  +\sum_{n\geq1}p_n\left[ q x_n + e^{\beta\omega} (1-q) (1-x_n)  \right]
\ee
Subtracting $q=q\sum_{n\geq0} p_n$ from both sides gives
\bea
\Delta q&=&(p_0-p_0)q  +\sum_{n\geq1}p_n\left[ q (x_n-1) + e^{\beta\omega} (1-q) (1-x_n)  \right]
\nonumber \\
&=& -\sum_{n\geq1}p_n(1-x_n)\left[ q  - e^{\beta\omega} (1-q)   \right]
\,,
\eea
implying that $\B$ is the excited to ground state occupation ratio in the stationary state.

\section{Energy flow in the diffusion limit \label{sec:energyflow}\labeld{sec:energyflow}}

\subsection{Energy flow in S}

At time $t$, we call
\be
E_S(t)=\Tr\left(\rho_S(t)H_S\right)=E\rho_{S,11}(t)
\,.
\ee
Then in the diffusion limit the flow of energy of S is
\be
\frac{dE_S}{dt} = -2\Gamma E e^{-2\Gamma t} \left[\rho_{S,11}(0)-\rho_{S,11}(\infty)\right]
\ee
which goes to zero.

\subsection{Flow of energy in A}

At time $t$ one has
\be
\frac{dE_A}{dt} = -\frac12\bar\gamma^2 \Tr\left(H_A[V,[V,\rho_S(t)\otimes\rho_A]] \right)
\ee
where $\rho_A$ is the thermal state of the bath, or after explicit calculation when $t\to+\infty$
\be
\frac{dE_A}{dt} \to \frac{-\frac12\bar\gamma^2}  {\sum_k |g_k|^2\coth\left(\frac{\beta\omega(k)}{2}\right)}
   \sum_{k,\ell}\frac
   {|g_k|^2|g_\ell|^2\left(\omega(k)-\omega(\ell)\right)\left(e^{-\beta\omega(k)}-e^{-\beta\omega(\ell)} \right)}
   {\left(1-e^{-\beta\omega(k)}\right)\left(1-e^{-\beta\omega(\ell)}\right)}
\ee
In particular, $\frac{dE_A}{dt}>0$ for $t\to+\infty$ unless the bath has only one frequency (tuned or detuned). In that case $\frac{dE_A}{dt}\to0$\@.

Therefore for a multi-frequency bath A, the bath will absorb energy at a constant rate, the energy coming from whatever external force is inducing the intermittent contact.

\subsection{Flow of interaction energy}

Each time the bath and system are disconnected the potential energy of interaction is canceled; thus
\be
\frac{dE_I}{dt}=-\frac{dE_A}{dt}-\frac{dE_S}{dt} \to \frac{dE_A}{dt}
\ee
where the last quantity represent the asymptotic value when S has reached its stationary state. Note that this depends on the initial interaction energy being zero (\cite{note:freecoupling}) as well as on the conservation of energy during the coupling interval. The asymptotic interaction energy flow, $\frac{dE_A}{dt}$, is negative if the bath has more than one frequency and results from the coupling/decoupling process at each time step. As indicated, it corresponds to the work on the system S provided by an external source.

\subsection{Spin boson energy flows}

We mention corresponding results for the SB interaction. As for RW, $\frac{dE_S}{dt}\to0$ exponentially. The flow of energy to the bath is
\be
\frac{dE_A}{dt}=-\bar\gamma^2\sum_k |g_k|^2\omega(k)<0
\,.
\ee
Once again, a constant flow of work must be provided by an external source during the coupling/decoupling process.


\begin{thebibliography}{12}
\bibitem[{not({\natexlab{c}})}]{note:kurizki}
\bibinfo{note}{For the of case small-$\Delta t$-refreshing, our work is similar
  to that of Ref.\ \cite{erez}. For their purposes the RW interaction is
  inappropriate and using non-Markovian effects interesting thermodynamical
  control is obtained. Our focus, however, is on RW, with SB serving as a foil.
  An analogous model for the study of the Quantum Zeno (and ``anti-Zeno'')
  Effect is \cite{lizuain}. In their model the bath is not refreshed to its
  original state, which need not be thermal.}
\end{thebibliography}

\end{document}